# A STRUCTURED APPROACH FOR THE IMPLEMENTATION OF DISTRIBUTED MANUFACTURING SIMULATION

Sameh M. Saad, Terence Perera and Ruwan Wickramarachchi
School of Engineering
Sheffield Hallam University
Sheffield, S1 1WB, United Kingdom

**ABSTRACT**

Manufacturing has been changing from a mainly in-house effort to a distributed style in order to meet new challenges owing to globalization of markets and world-wide competition. Distributed simulation provides an attractive solution to construct cross enterprise simulations to evaluate the viability of the proposed distributed manufacturing enterprises. However, due to its complexity and high cost distributed simulation failed to gain a wide acceptance from industrial users. The main objective of this paper is to address these issues and present a new structured approach to implement distributed simulation with cost effective and easy to implementable tools. A simplified approach for model partitioning for distributed simulation is also included in the proposed approach. The implementation of distributed manufacturing simulation is illustrated with Arena, Microsoft Message Queue (MSMQ) and Visual Basic for Applications (VBA).

**INTRODUCTION**

In today's highly competitive world, manufacturing enterprises are confronted with growing competition, the evolution of new markets, more and more sophisticated consumer demand, and increasingly complex global political and economic scenarios. In order to lower the costs, increase profits, reduce product development times, enhance products, and react to environmental changes more positively manufacturing enterprises are moving towards open architectures for integrating their activities with those of their suppliers, customers and partners. In manufacturing, companies may form strategic partnerships for outsourcing some of their operational activities, share resources or joint development of products and services etc., leading to formation of virtual manufacturing enterprises which operate in distributed manufacturing environment. To facilitate the creation of virtual manufacturing enterprises, potential partners must be quickly able to evaluate whether it will be profitable for them to participate in the proposed enterprise. Simulation provides a capability to conduct experiments rapidly to predict and evaluate the results of manufacturing decisions (McLean and Leong, 2001).

As Law and McComas (1998) noted manufacturing is one of the largest application areas of simulation, with the first uses dating back to at least early 1960s. However, traditional sequential simulation alone may not sufficient to simulate highly complex Distributed Manufacturing Enterprises (DME). In such situations, distributed simulation provides a promising alternative to construct cross enterprise simulations. The use of distributed simulation allows each partner to hide any proprietary information in the implementation of the individual simulation, simulate multiple manufacturing systems at different degrees of abstraction levels, link simulation models built using different simulation software, to take advantage of additional computing power, simultaneous access to executing simulation models for users in different locations, reuse of existing simulation modes with little modifications etc. (Venkateswaran et al., 2001; McLean and Riddick, 2000; Gan et al., 2000; Taylor et al., 2001). However, Peng and Chen (1996) noted that as a technique, parallel and distributed simulation is not very successful in manufacturing. Most of the distributed manufacturing simulations developed were implemented with either simulation languages or general purpose programming languages such as C++ and Java. This calls expertise not only in distributed simulation but also in programming too. Moreover, general business community is not very receptive towards distributed simulation due to its complexity, long development times, involvement of step learning curves, high costs etc.

Another important issue needs to be addressed when designing a distributed simulation is partitioning of the simulation model into sub-models or logical processes (LPs). Efficiency and effectiveness of a distributed simulation system depends on partitioning of the system. Some of the existing approaches require executing of the whole model sequentially in order to collect data before partitioning and mapping carried out based on data collected. However, a simulation is executed in distributed manner because its inability to run sequentially due to size, complexity, requirements for more computing resources, or need to run geographically distributed manner etc. This creates a dilemma for users especially in business organizations, who intend to design parallel and distributed simulations.

The objective of this paper is to present a new simplified approach to implement distributed manufacturing simulation (DMS). It includes a simplified approach to model partitioning and mapping, and simulation model development processes for DMS. Instead of implementing the distributed simulation with programming languages, we are proposing to develop the system using commercial simulation software.

**BACKGROUND**

Distributed simulation combines distributed computing technologies with traditional sequential simulation techniques. In recent years, popularity of distributed computing applications increased due to proliferation of inexpensive and powerful workstations, improvements in networking technologies, low cost equipment and incremental scalability. Hence, the use of network of workstations has been evolving into a popular and effective platform for distributed simulation. However, low communication speeds, shortage of network bandwidth and the ever increasing demand for network resources may result slowing down the execution speed of the distributed simulation model. Although the networked workstations are slower than dedicated machines, they may be fast enough and may require much less specialist expertise to put them to use with a fraction of a cost of the price needed for a dedicated parallel processing computer (Cassel and Pidd, 2001).

Throughout the century, the world of manufacturing has changed from a mainly in-house effort to a distributed style of manufacturing. As the term distributed manufacturing implies, DMEs which also known as virtual manufacturing enterprises are ephemeral organizations in which several companies collaborate to produce a single product or product line (Venkateswaran et al., 2001). Participating in this type of collaboration allow partner organizations to use their knowledge, resources and in particular manufacturing expertise to take advantage of new business opportunities and/or gain a competitive advantage that are on a larger scale than an individual partner could handle alone.

**PROPOSED APPROACH**

**Modeling and Model Partitioning for Distributed Manufacturing Simulation**

The degree to which the simulation results are able to characterize the system under study is directly related to the degree the simulation model characterizes the system (Luna, 1993). In order to understand the problems, requirements and perhaps alternative solutions for many systems especially complex and large ones, it is desirable to build a conceptual model before transforming it into a computer simulation model. Conceptual model is a simulation developer's way of translating modeling requirements (ie. What to be represented by simulation) into detailed design framework (ie. How it is to be done), from which the software that will make up the simulation can be built (Pace, 1999). Furthermore conceptual model is the ultimate expression of the system functionality and should be the basis for testing and verification and validation procedures (Haddix, 2001).

A conceptual model developed with an appropriate modeling approach and modeling tool facilitates partitioning of the DME model into LPs. Modeling approaches specify the way models are to be developed while Modeling tools provide a standard means of describing and analyzing systems. Hierarchical modeling approach was selected since it provides a way of managing large scale complex systems by considering them as a collection of sub-systems (Kiran, 1998). In a distributed simulation system these sub systems are represented by simulation models that are independently created, modified and saved. Pidd and Castro (1998) also noted that many large systems are inherently hierarchical.

IDEF0 was chosen as the modeling technique for the proposed approach, and it has been widely used in industry due to its user-friendliness, computer support, rigor and conciseness, and well documented rules and procedures (Pandya, 1995; Kateel et al., 1996). Number of authors including Cheng-Leong et al. (1999), Cheng-Leong (1999), Whiteman et al. (1997), Rensburg and Zwemstra (1995) have highlighted the usefulness of IDEF0 as a model representation technique in simulation. Another benefit of using IDFE0 with commercial simulation software is that IDEF0 structure of the model can easily be transformed into simulation model. Figure 1 shows a part of simulation model developed by Arena for an IDEF0 model.

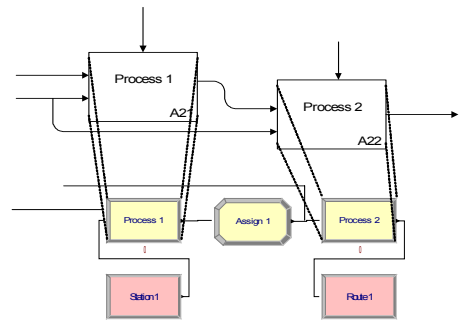

Figure 1: IDEF0 diagram and Arena simulation model

With hierarchical modeling approach and IDEF0 technique, LPs that can function independently could be identified based on interactions between different sections. In the IDEF0 model these interactions are represented by number of lines between boxes that represent different sections of the enterprise.

Once the LPs are identified, they could be validated to make sure that LPs represent individual entities of the enterprise, and the entire enterprise when considered together. The validated LPs could be assigned (mapped) into workstations in a computer network before converting them into computer simulation models and execute as a distributed simulation. Figure 2 shows the proposed approach for modeling, model partitioning and mapping for the distributed manufacturing simulation.

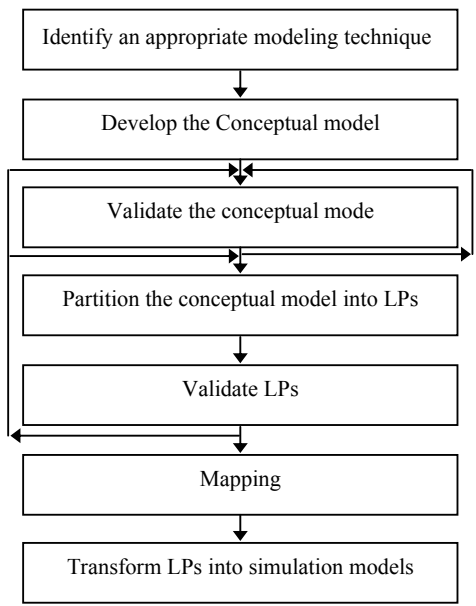

Figure 2: Proposed approach for modeling, model partitioning and mapping

The main difference between existing approaches and the proposed approach is the stage of partitioning carried out in the simulation methodology. According to most of the current approaches partitioning is done only after the system is converted into a computer program using algorithms in order to minimize the communication overheads and optimize the load balance. To simplify the distributed simulation development process it is proposed to partition the conceptual model into LPs and assign them into workstations before transforming LPs into computer simulation models. It is also assumed that networked workstations are freely available to assign LPs and only one LP is mapped into a workstation.

**Development of the Distributed Manufacturing Simulation**

Distributed Simulations pose unique synchronization constraints due to their underlying sense of time. When the simulation state can be simultaneously changed by different processes, actions by one process can affect actions of another (Nicol, 1993). In order to make sure that each LP processes arriving messages in their timestamped order and not in their real time arriving order, individual simulation models needed to be synchronized. This requirement is referred to as local causality constraint (Fujimoto, 1990). Optimistic protocols are implemented by saving simulation state at different points of time and rolling back to a previous time point if local causality constraint is violated. If a programming language is used to develop the simulation, then state saving mechanism can be integrated into the distributed simulation engine itself. Since commercial simulation packages generally do not allow saving simulation state at different time points and rolling back to previous time points, it was decided to employ conservative protocol to synchronize the distributed manufacturing simulation. Conservative approaches strictly impose the local causality constraint and guarantee that each model will only process events in non-decreasing timestamp order. Determining a value for lookahead is one of the most important and difficult aspect of conservative protocol. However, it was assumed that minimum-processing times (which can be used as lookahead values) for LPs can be calculated. An approximate synchronization mechanism, especially suitable for distributed manufacturing applications has been proposed by Saad et al. (2003).

In order to synchronize and pass parameters, simulation models need to communicate with each other. Communication methods provided by operating systems often require complex programming. In a distributed simulation, middleware provides simple and reliable solution for this problem. Middleware is a class of software designed to help manage the complexity and heterogeneity inherent in distributed system. It contains a set of enabling services which allow multiple processes running on one or more computers to interact across a network. Analysis of past literature reveals number of attempts to simulate distributed manufacturing systems and supply chains using tools such as HLA, CORBA and GRIDS (see Venkateswaran et al., 2001, Taylor et al., 2001; Gan et al., 2000; McLean and Riddick, 2000). For the proposed approach, Microsoft Message Queue (MSMQ), a Message Oriented Middleware was selected to link simulation models.

As MSMQ is integrated into newer versions of Windows operating systems and available as an additional component for Windows NT, 98 and 98, it provides a cost effective solution for message passing. MSMQ interacts with simulation model through an application program interface (API). Arena simulation software was used in this study as commercial simulation software to demonstrate the implementation. However, other commercial simulation software such as Automod, Promodel, Witness etc. can also be used for this purpose. Both Arena and MSMQ support Visual Basic for Applications (VBA) and C++. Since, programming of Arena with VBA is more straightforward than with C++, it was decided to use VBA to develop the API. VBA also offers a programming environment similar to popular Visual Basic programming language.

API developed for MSMQ could send messages containing parameters obtained from simulation model to a queue in the same computer or directly to another remote computer. API that resides in the remote computer extracts these messages from the queue and passes the parameters to the simulation model (Figure 3).

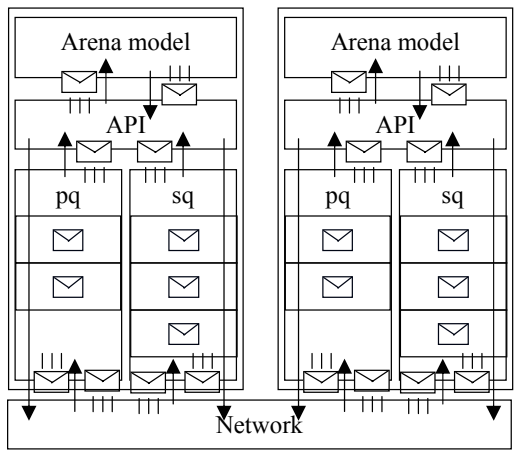

Figure 3: MSMQ, API and Arena

## ILLUSTRATION

In order to illustrate the development of distributed simulation with Arena, MSMQ and VBA, following brief case study is presented.

A, B and C firms are proposing to form a distributed manufacturing enterprise (Figure 4) to produce a new product called XYZ. Firm A is to produce and process parts X and Z. Part Z is sent to Firm C and part X is sent to Firm B which posses an expensive processing unit for further processing. Firm B is also to produce part Y and assemble Parts X and Y together to form component XY which is then sent to Firm C for further processing and final assembly. At Firm C, component XY and part Z are to be further processed and assembled together to produce product XYZ. In addition to processing of parts X, Y, Z, component XY and product XYZ, three firms also produce their own products independently. Parts are to be passed in batches of 1000s and transfer time from one firm to another firm was assumed as 10 hours. Before committing on the DME, firms want to evaluate the feasibility in terms of capacity utilization and how the proposed venture affects their existing operations. As firms are reluctant to pass information of their processes to other firms, it was agreed to develop 3 models separately and run them in a distributed simulation environment.

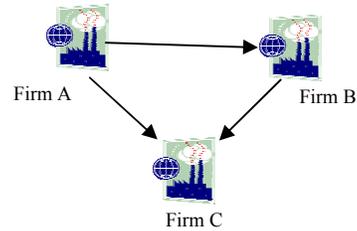

Figure 4: A model for distributed manufacturing

Three Arena simulation models were independently developed, verified and validated for firms A, B and C. Entities created within models were used instead of inputs from other models for B and C. Then B and C were modified by replacing the 'Create module' with a 'Create block' and adding a 'VBA block' just before the 'Dispose module' (Figures 5 and 6). 'Create block' can be used to release entities into the model which created by API of the model to represent output from A and/ or B. Two MSMQ queues were created in each workstation, one to accept inputs (pq) and the other to synchronize (sq) the distributed simulation. Once batch of 1000 units were processed at A or B, the batch goes through a VBA block and API written in VBA sends a message to destination model (Figure 6). When a message is reached its destination queue, it is processed automatically with built-in 'qevent' event. Once a message comes to 'pq', 'qevent' creates an entity and schedules to release it after 'transfer time' at 'Create block'. The 'Separate module' adds additional 999 units to make a batch of 1000 which was passed as output from previous model (Figure 5). At model C, output from A and B can be identified by message label. Figures 7 and 8 show sample code written for VBA block and 'qevent' respectively.

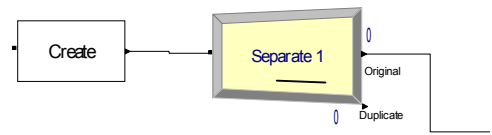

Figure 5: Adding output received from other models as input

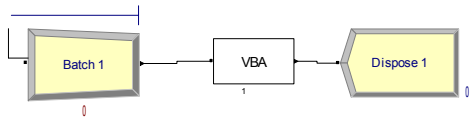

Figure 6: Passing output to destination model VBA block

```
Private Sub VBA_Block_1_Fire()
  Dim qinfo As MSMQQueueInfo
  Set qinfo = New MSMQQueueInfo
  qinfo.FormatName = "DIRECT = OS:ENG-4l30-10
          \private$\mbq"
  Dim qQueue As MSMQQueue
  Set qQueue = qinfo.Open(MQ_SEND_ACCESS,
          MQ_DENY_NONE)
  Dim qMsg As MSMQMessage
  Set qMsg = New MSMQMessage
  qMsg.Label = "A"
  qMsg.Body = "1000"
  qMsg.Send qQueue
  qQueue.Close
End Sub
```

Figure 7: Sample code of VBA block

```
Sub qEvent_Arrived(ByVal Queue As Object, ByVal
cursor As Long)
  Dim vEntityIndex As Long
  Dim vPictureIndex As Long
  Dim aQueue As MSMQQueue
  Set aQueue = Queue
  Dim qMsg As MSMQMessage
  Set qMsg = aQueue.Receive(, , , 0)

  vPictureIndex =
ThisDocument.Model.SIMAN.SymbolNumber
          ("Picture.Package")
  vEntityIndex = ThisDocument.Model.SIMAN
          .EntityCreate
  Call ThisDocument.Model.SIMAN.EntitySetPicture
          (vEntityIndex, vPictureIndex)

  If qMsg.Label = "A" Then
    Call ThisDocument.Model.SIMAN
          .EntitySendToBlockLabel(vEntityIndex
          , 10, "CreateBlockA")
  Else
    Call ThisDocument.Model.SIMAN
          .EntitySendToBlockLabel(vEntityIndex,
          10, "CreateBlockB")
  End If

  aQueue.EnableNotification qEvent
End Sub
```

Figure 8: Sample code to create and schedule an entity to represent input (of model C)

## DISCUSSION

This paper presented a simplified approach to implement a distributed manufacturing simulation. Although a distributed manufacturing application was used to illustrate the implementation, the proposed approach may be able to use in other application areas too. The main benefit of this research is the simplified approach employed when developing the distributed simulation models using commercial simulation software and, connecting and running them in distributed environment with MSMQ and VBA. Not only Arena, but also other simulation software packages such as Promodel, Automod and Witness can be used to implement the simulation. Furthermore, simulation models developed with different simulation software can be connected together and run as a distributed simulation as long as they support either VBA or C++. Cost involved can be kept low as no additional costs involved with middleware and application program interface (API), provided workstations are running on a windows operating system. The proposed approach also encourages reusability of existing simulation models. Existing simulation models developed for traditional sequential simulation require only minor modifications to adopt for a distributed simulation.

It is expected that this simplified approach may address criticisms made against distributed simulation because of its complexity to develop and implement, higher costs involved, need for more expertise etc. Working of the simulation model can be animated easily as commercial simulation software is used for implementation. Animation may play a very effective role in convincing the benefits of simulation to non simulation users such as managers, workers etc.

Main shortcoming of the proposed approach is the sacrifice made in performance of the distributed simulation mainly in terms of speedup. It may also not be feasible to employ the proposed approach to implement highly complex systems such as telecommunications systems, computer networks, logic circuits etc. However, applications which are not executed in distributed manner only to gain speedups, and applications that specifically require executing in distributed simulation environment are ideally fit for the approach we presented in this paper.

## CONCLUSIONS

The proposed approach addressed some of the criticisms leveled against distributed simulation with cost effective and simplified implementation approach for distributed manufacturing simulation.

Performance comparisons between distributed simulation implemented using proposed approach and conventional approaches may provide an opportunity to fully understand the benefits and the shortcoming of our work. Unlike sequential simulation, output from a distributed simulation can be obtained for individual models as well as for the whole system under investigation. Since outcome of the simulation effort depends on the output of the simulation, it may worth investigating strategies to identify and generate output locally at individual models and for the entire distributed simulation system.

## AUTHOR BIOGRAPHIES

**SAMEH M. SAAD** (BSc, MSc, PhD, CEng, MIEE, ILTM) is a Reader in Advanced Manufacturing Systems and Enterprise modeling and management and Postgraduate Course Leader at the Systems and Enterprise Engineering Division, one of the three divisions in the School of Engineering, Sheffield Hallam University, United Kingdom. Dr Saad's research interests revolve around aspects of design and analysis of manufacturing systems, production planning and control, systems integration, reconfigurable manufacturing systems, manufacturing responsiveness, enterprise modeling and management and next generation of manufacturing systems. He has published over 55 articles in various national and international academic journals and conferences. His contact email address is <s.saad@shu.ac.uk>

**TERRENCE PERERA** (BSc, PhD) is Professor and Head of Enterprise and Systems Engineering at School of Engineering, Sheffield Hallam University. He also leads the Systems Modeling and Integration Research Group. His current research interests include the implementation, integration and practice of virtual modeling tools within all industrial sectors. His email address is <t.d.perera@shu.ac.uk>

**RUWAN WICKRAMARACHCHI** is a PhD student at Sheffield Hallam University, United kingdom. He received his MPil and BSc degrees from University of Cambridge, United Kingdom and University of Kelaniya, Sri Lanka respectively. His main research interest focused on distributed enterprise simulation with emphasis on distributed manufacturing applications. He can be contacted by <w.ruwan@shu.ac.uk>